\documentclass[11pt,letterpaper,oneside,pdftex]{article}
\pdfoutput=1
\usepackage[utf8]{inputenc}
\usepackage[T1]{fontenc}
\usepackage{amsmath}
\usepackage{amsfonts}
\usepackage{amssymb}
\usepackage{graphicx}
\usepackage[width=16.20cm, height=22.50cm]{geometry}
\usepackage{fouriernc}
\usepackage{soul}

\usepackage[margin=1cm,font=small]{caption}
\usepackage[numbers,square,comma,sort&compress]{natbib}
\usepackage[pdftex,hyperref,svgnames]{xcolor}
\usepackage[pdftex,bookmarksnumbered=true,breaklinks=true]{hyperref}
\makeatletter
\AtBeginDocument{
	\hypersetup{
		pdftitle = {\@title},
		pdfauthor = {\@author}
	}
}
\makeatother
\makeatletter

\@addtoreset{equation}{section}
\makeatother

\newcommand{\vv}[1]{\mathbf{#1}}
\title{\texorpdfstring{\begin{flushright}
			{\small LA-UR-22-29552}
		\end{flushright}\vspace{2em}}{}%
Predicting electrical conductivity in bi-metal composites}
\author{Daniel N. Blaschke, John S. Carpenter, and Abigail Hunter}
\date{Oct. 15, 2024}

\graphicspath{{./}{./figures/}}
\begin{document}
\maketitle

\thispagestyle{empty}
\begin{center}
	\vspace{-0.3cm}
	Los Alamos National Laboratory, Los Alamos, NM, 87545, USA
	\\[0.5cm]
	{\ttfamily{E-mail: dblaschke@lanl.gov, carpenter@lanl.gov, ahunter@lanl.gov}}
\end{center}

% \vspace{1.5em}

\begin{abstract}
Generating high magnetic fields requires materials with not only high electric conductivity, but also good strength properties in order to withstand the necessarily strong Lorentz forces.
A number of bi-metal composites, most notably Cu/Nb, are considered to be good candidates for this purpose.
Here, we generalize our previous work on Cu/Nb in order to predict, from theory, the dependence of electric conductivity on the microstructure and volume fraction of the less conductive component for a number of other bi-metal composites.
Together with information on strength properties (taken from previous literature), the conductivity information we provide in this work can help to identify new promising candidate materials (such as Cu/Nb, Cu/Ag, Cu/W, \ldots) for magnet applications with the highest achievable field strengths.
\end{abstract}

\vspace{1cm}
% \newpage
\tableofcontents
\newpage

\section{Introduction}
\label{sec:intro}
%%%%%%%%%%%%%%%%%%%%%%%

Studying fundamental questions and behaviors (such as phase transformations) in a wide range of materials systems such as semi- and super-conductors, quantum matter and thin films, often requires the creation of ultra-high magnetic fields (>80T).
The current world record for pulsed magnetic fields is 100 T, which was set about a decade ago \cite{Feder:2012}.
While there is high demand for the use of such ultra-high magnetic field resources, access can still be limited for multiple reasons, one of them being the material properties of the conductive wire used within the magnet \cite{Nguyen:2016}.
Currently, the 100 T pulsed field magnetic at the National High Magnetic Field Laboratory utilizes a Cu/Nb nano-composite wire for this purpose \cite{Nguyen:2016}.
This wire is manufactured using an accumulative drawing and bonding (ADB) method that makes it difficult to produce consistent material properties from batch-to-batch and over the relatively long lengths required in the magnet \cite{Nguyen:2016}.
Variation in material properties can cause experiments to either out-right fail, or have sub-par performance, resulting in costly delays and limitations in the availability of this resource. 

These conductive wires are required to have a unique set of material properties. Particularly, in order to withstand the necessarily high Lorentz forces, the conductive winding wire utilized in these applications must have both high conductivity as well as exceptional strength properties  \cite{Foner:1989,Campbell:1995,Embury:1998,Han:1999,Freudenberger:2002}.
For example, the current Cu/Nb wire has an ultimate tensile strength (UTS) of $\sim$1 GPa and an electrical conductivity of $\sim$70\% IACS (International Annealed Copper Standard) at room temperature \cite{Han:2002,Liang:2010}.
Furthermore, the wire needs to be fairly ductile as it is wound into coils to be used in the magnet \cite{Nguyen:2016}.
Finally, these ultra-high pulsed field magnets operate at low temperatures ($\sim77$--400\,K \cite{Nguyen:2016}), so the material properties of the wires must be maintained through a range of temperatures.
For example, at 77K, Cu/Nb can exhibit very high electrical conductivities, $\sim$300\% IACS and an UTS ranging from $\sim$1-1.4 GPa \cite{Han:2002,Liang:2010}.
New composite materials with improved material properties regarding electrical conductivity and strength help to open the door for even higher pulsed magnetic fields, which would enable new experimental possibilities for several different classes of materials.

Promising material candidates for this purpose that so far have been identified include Cu/Nb \cite{Dupouy:1995,Foner:1989,Pantsyrnyi:2001,Han:2002}, Cu/stainless steel \cite{Dupouy:1995, Pantsyrnyi:2001}, Cu/Cr \cite{Dobatkin:2015}, Cu/W \cite{Dong:2020,Han:2024}, Cu/Ta \cite{Zeng:2016}, and Cu/Ag \cite{Campbell:1995,Sakai:1991,Zhao:2016}.
All of these examples are two-phase composites which are fabricated using severe plastic deformation (SPD) methods such as ADB or accumulative roll bonding (ARB).
Because it is currently used in ultra-high pulsed field magnets, the majority of previous literature has been focused on understanding and optimizing Cu/Nb.
Thus, the material properties of Cu/Nb naturally become the baseline performance on which the community aims to improve upon.
As mentioned, several other candidate materials have been investigated, and with the recent work on multi-principal element and high-entropy alloys, one can imagine a vast design space in which to explore.

The focus of this current work is on bi-metal composites, and as such we continue our previous study \cite{Blaschke:2022CuNb}, of understanding and predicting electric conductivity based on the microstructure.
Together with a separate study of material strength (not included here), this endeavor will help to identify new materials with even more optimized conductivity/strength properties for magnet applications.
In particular, we generalize our previous theoretical study of Cu/Nb to a number of other bi-metals and identify the volume fraction the second (less conductive) component should not exceed in order to maintain good conductivity.
A related recent numerical study of Cu/Nb composites can be found in Ref. \cite{Shiraiwa:2024}.

The paper is organized as follows:
We start in Section \ref{sec:model} by reviewing the modeling techniques, i.e. the phase field framework and the models predicting the contributions of the various microstructures to electric resistivity.
We then proceed in Section \ref{sec:results} with presenting our simulation results for electric conductivity of a number of bi-metals.
Strength properties of those bi-metals are assembled from the literature, and we give an overview over promising candidates of strong and conductive bi-metals in Section \ref{sec:strength}.

\section{The model}
\label{sec:model}
%%%%%%%%%%%%%%%%%%%%%%%

Like in our previous work \cite{Blaschke:2022CuNb} and following \cite{Jin:2013}, we consider local charge density $\rho_e$ as our order parameter within a phase field approach and (within our simulations) must apply an external electric field $E_\text{ex}$.
The local electric field $\vv{E}$ is generated by the spatial distribution of charge density $\rho_e$ and the externally applied electric field, i.e.
\begin{align}
\vv{E}(\vv{x}) &= \vv{E}^\text{ex} - \frac{i}{\varepsilon_0}\int \frac{d^3k}{(2\pi)^3} \frac{\tilde\rho_e(\vv{k})}{k^2}\vv{k}e^{i\vv{k}\vv{x}}
\,,\label{eq:Efieldlocal}
\end{align}
where $\tilde{\rho}_e$ is the Fourier transform of $\rho_e$.
The local current density $\vv{j}$ is then given (via the microscopic version of Ohm's law) by
\begin{align}
j_i &= \sigma_{ij} E_j(\vv{x})
= \sigma_{ij} E_j^\text{ex}  - \frac{i \sigma_{ij}}{\varepsilon_0}\int \frac{d^3k}{(2\pi)^3} \frac{{k}_j}{k^2}\tilde\rho_e(\vv{k}) e^{i\vv{k}\vv{x}}
\,, \label{eq:currentlocal}
\end{align}
where conductivity $\sigma$ may take a tensorial form to account for anisotropy.
For simplicity, we however assume an isotropic approximation for $\sigma_{ij}\approx \sigma\delta_{ij}$.

The law of charge conservation, $\dot\rho_e=-\nabla\vv{j}$, is related to the Ginzburg-Landau equations within a phase-field formulation.
Hence, evolving this equation will give us a final charge density distribution, which via Eqs. \eqref{eq:Efieldlocal} and \eqref{eq:currentlocal} yields a current distribution $\vv{j\vv({x})}$, i.e.
\begin{align}
\dot\rho_e&=-\partial_i\left(\sigma E_i(\vv{x})\right)
=-\partial_i\left(\sigma  {E}_i^\text{ex} \right)
-\frac{\sigma}{\varepsilon_0}\int \frac{d^3k}{(2\pi)^3} \tilde\rho_e(\vv{k}) e^{i\vv{k}\vv{x}}
+\frac{i}{\varepsilon_0}\left(\partial_i\sigma\right)\int \frac{d^3k}{(2\pi)^3} \frac{{k}_i}{k^2}\tilde\rho_e(\vv{k}) e^{i\vv{k}\vv{x}}
\,. \label{eq:LGeqConduct}
\end{align}
The experimentally measured macroscopic conductivity is then determined by spatially averaging $\vv{j}$ and the macroscopic version of Ohm's law:
\begin{align}
	\langle j_i\rangle &= \sigma^\text{eff} E_i^\text{ex}
	\,.
\end{align}
In order to model local conductivity $\sigma$, we assume Matthiessen's rule \cite{Matthiessen:1864} holds, i.e. electric resistivity $\rho=1/\sigma$ is the sum of bulk resistivity $\rho_0(T)$ and additional contributions to the bi-metal interfaces ($\rho_i$), grain boundaries ($\rho_\text{gb}$), and average dislocation density $\rho_d$.
We neglect sub-leading contributions from vacancies, interstitial atoms, and other types of defects, assuming they are not prevalent in our samples.
The three contributions to $\rho$ can be modeled as follows \cite{Blaschke:2022CuNb,Ding:2021,Tian:2014,Dingle:1950,Fuchs:1938}:

The probability of scattering electrons at a bi-metal interface is parameterized by model parameter $p\in[0,1]$ and its value depends on the ``roughness'' of the bi-metal interface with typical values being in the vicinity of 0.5 \cite{Tian:2014}.
The contribution to resistivity in the two phases of the bi-metal on either side of the interface stemming from such scattering events an be estimated as\footnote{
Note that after careful consideration we realized the additional factor ${2V_f}/{(1-V_f)}$ depending on the volume fraction $V_f$ of the second phase (which is 1 for $V_f=1/3$) present in \cite{Blaschke:2022CuNb} should not be included.
Also, we re-calibrated the fraction of layers $\ge100$nm containing 2 grains (30\% instead of 25\%) as well as the standard deviation of the Gaussian distribution (50nm instead of 40nm); see the details given below in Sec. \ref{sec:results}.}
\begin{align}
\rho_\text{if} &= \rho_{0}(T)\left[\frac{3}{8}(1-p)\lambda_0(T)\frac{1}{d_0}
% \Bigg\{\begin{array}{c}
% 1\\ \frac{2V_f}{1-V_f}
% \end{array}
\right]
\,, \label{eq:ifscattering}
\end{align}
where
% $V_f$ denotes the volume fraction of the second phase, 
$d_0$ denotes the layer thickness, and $\lambda_0$ is the electron mean free path.
Both $d_0$ and $\lambda_0$ can be different on either side of the interface.
Note that $\rho_\text{if}$ increases significantly as the layer thickness shrinks to values competing with the electron mean free path.

Likewise, grain boundary scattering becomes important when grain sizes are small so that they are comparable to $\lambda_0$.
The according contribution to electric resistivity is estimated from the following expression:
\begin{align}
\rho_\text{gb} &= \rho_0(T)\left[\left(1-\frac32\alpha+3\alpha^2-3\alpha^3\ln\left(1+1/\alpha\right)\right)^{-1} - 1\right]\,,
\nonumber\\
\alpha &= \frac{\lambda_0(T)}{d}\frac{R}{1-R}
\,, \label{eq:gbscattering}
\end{align}
where the grain boundary reflection coefficient $R\in[0,1]$ depends on the size and shape of the single crystal grains.
Grains in an ARB material with thin layers are typically long and flat \cite{Zhang:2022}, so that $d$ (the grain ``size'') is taken to mean the grain thickness perpendicular to the layer orientation.

Dislocation density, finally is parameterized by $\rho_d=R_dN_d$, where the ``scattering power'' $R_d$ is typically of the order of $10^{-25}$ $\Omega\,$m$^3$ \cite{Brown:1977} so that dislocation densities ($N_d$) below $10^{16}$m$^{-2}$ lead to negligible contributions to electric resistivity.
Experiments with Cu/Nb have shown \cite{Blaschke:2022CuNb} that typical dislocation densities are below or close to $10^{15}$m$^{-2}$.
Here, we assume this value for $N_d$ in our simulations, even though its effect is in the sub-percent level and thus very small.

\section{Conductivity Results}
\label{sec:results}
%%%%%%%%%%%%%%%%%%%%%%%%

\begin{figure}[!htb]
\centering
\includegraphics[width=0.75\textwidth]{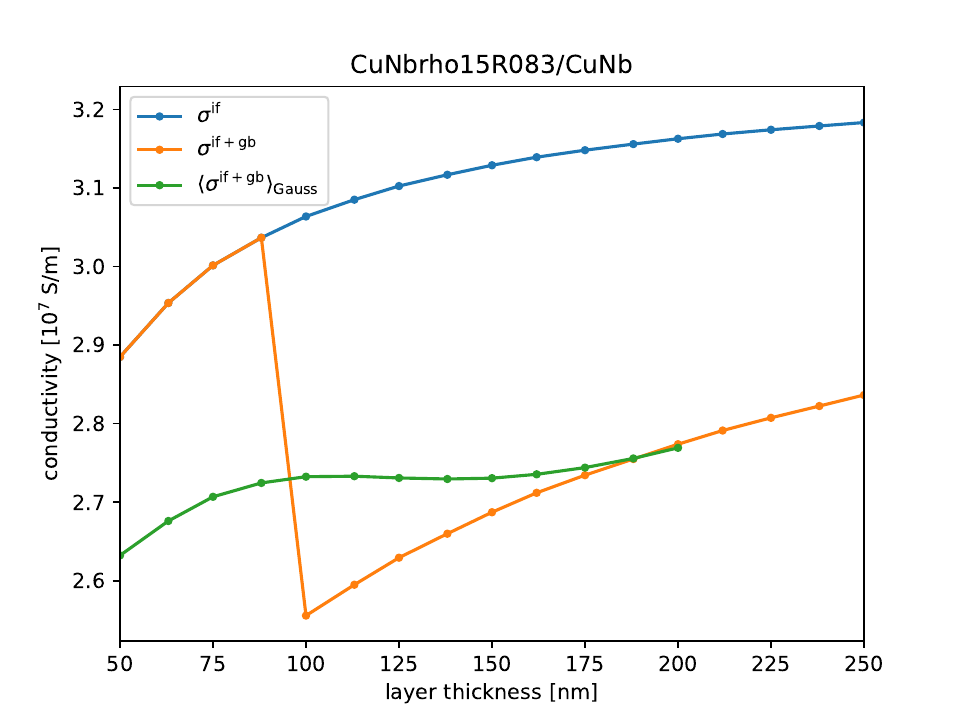}
\caption{We show electric conductivity as a function of layer thickness for a Cu/Nb composite.
The blue curve shows model results taking into account only the effect of interface scattering.
For the orange curve, we have assumed 30\% of the thicker layers ($\ge100$nm) have 2 grains across the layer thickness.
The green curve shows the effect of having a distribution of layer thicknesses and was computed using a Gaussian distribution of the results shown in orange with a standard deviation of 50nm.
This rough calibration is based on our previous work, Ref. \cite{Blaschke:2022CuNb}, using our own measured Cu/Nb data which exhibited a range of layer thicknesses within each sample.}
\label{fig:CuNb05Vf300T_if_gb_spread}
\end{figure}

\begin{figure}[!htb]
\centering
\includegraphics[width=0.75\textwidth]{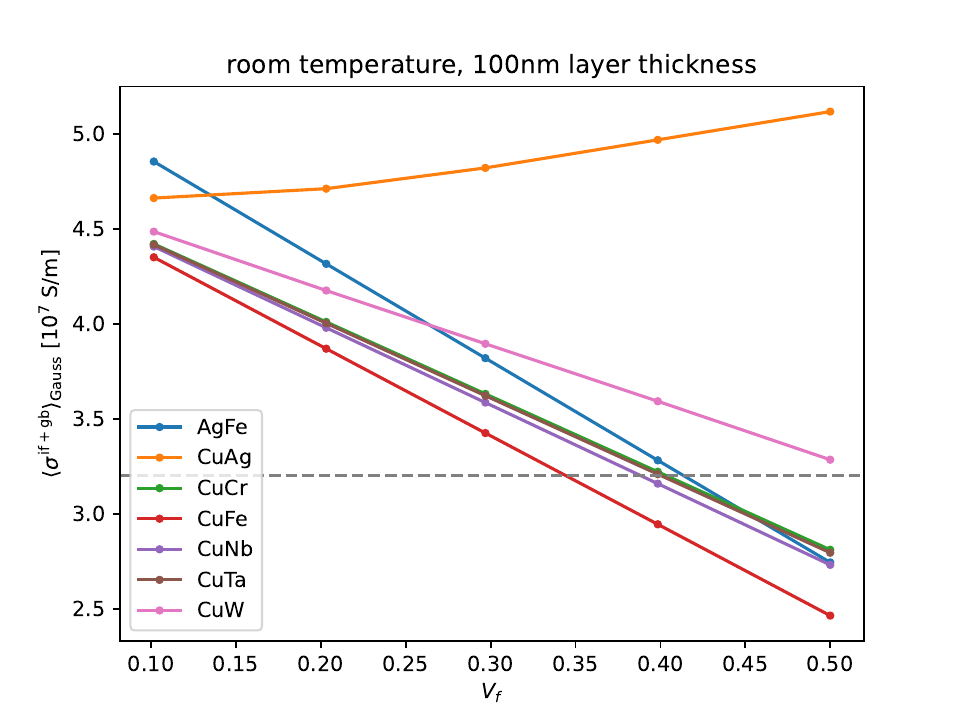}
\caption{We show electric conductivity as a function of volume fraction for a number of bi-metal composites at room temperature.
The  mean value within a Gaussian distribution of layer thicknesses with 50nm standard deviation for the first metal was 100nm.
%The average layer thickness of the first metal was 100nm, i.e. this was the mean value within a Gaussian distribution of layer thicknesses with 50nm standard deviation.
%The second metal's layer thickness equals that of the first metal only at $V_f=0.5$ and is typically thinner when the second metal's volume fraction $V_f<0.5$ (see e.g. \cite{Ding:2021,Carpenter:2023a});
30\% of the thicker layers were assumed to have two single crystal grains across the thickness.
The dashed gray line indicates the conductivity required in ultra-high magnetic field applications.}
\label{fig:RTsimul_Vf_wsp}
\end{figure}

\begin{figure}[!htb]
\centering
\includegraphics[width=0.75\textwidth]{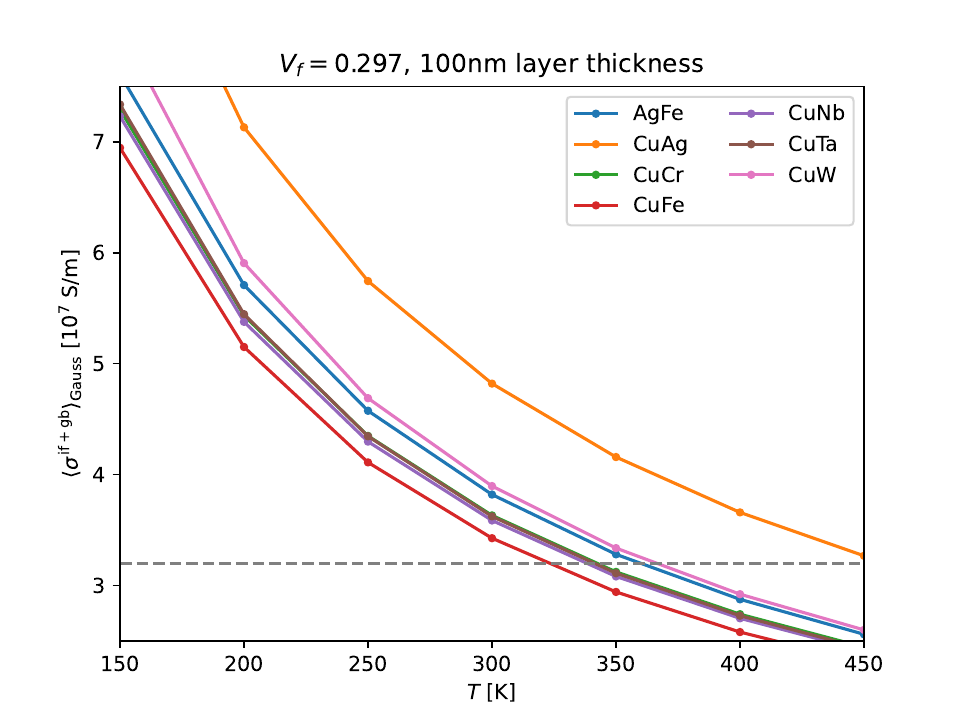}
\caption{We show electric conductivity as a function of temperature for a number of bi-metal composites.
The volume fraction of the second metal is $V_f\approx0.3$ in all of these simulations.
The average layer thickness was 100nm, i.e. this was the mean value within a Gaussian distribution of layer thicknesses with 50nm standard deviation.
30\% of the thicker layers were assumed to have two single crystal grains across the thickness.
The dashed gray line indicates the conductivity required in ultra-high magnetic field applications.}
\label{fig:Vf03simul_T_wsp}
\end{figure}

In our previous work, we showed that a typical Cu/Nb bi-metal exhibits a distribution in layer thickness, rather than equally thick layers, an effect that is even more pronounced when the volume fraction $V_f$ differs from 0.5 \cite{Blaschke:2022CuNb,Carpenter:2023a}.
Furthermore, for sufficiently thin layers, most of those layers have only one single crystal grain across the layer thickness and a smaller fraction of layers of thicknesses beyond 100nm exhibit 2, and less frequently more grains across the thickness.
Lacking details of the grain distribution, a rough first-order approximation, which led to good theoretical predictions for conductivity, was achieved by calibrating the fraction of layers thicker than 100nm to have 2 grains across the thickness \cite{Carpenter:2013}, and only one grain otherwise.
Combining this assumption with a Gaussian distribution in layer thickness (which could be changed if new data became available) led to good agreement with measured conductivity values for Cu/Nb with average layer thicknesses up to 150nm, as shown in our previous work \cite{Blaschke:2022CuNb}.
Figure \ref{fig:CuNb05Vf300T_if_gb_spread} illustrates these two steps:
The blue curve within this figure shows the theory predictions when only one grain is present across each layer thickness (re-calibrated in anticipation of Fig. \ref{fig:conduct_vs_strength_overview} below).
Adding two grains to 30\% of thicker grains leads to a drop in conductivity in those thicker layers, as shown in orange.
A Gaussian distribution over the layer thicknesses (calculated from the orange curve) yields the final prediction in green.

Assuming that the same ARB processing techniques leads to similar grain sizes, we now proceed to apply those same assumptions to predict conductivity for a number of other bi-metals of interest, in particular:
Ag/Fe, Cu/Ag, Cu/Cr, Cu/Fe, Cu/Nb, Cu/Ta, and Cu/W.
Note that more accurate predictions would require detailed knowledge of the grain distribution within each bi-metal, which would have to be measured.
Lacking this knowledge, the next best thing we can do presently is to use the same calibration and assumptions which led to good agreement with our Cu/Nb experiments in Ref. \cite{Blaschke:2022CuNb}.

In Figure \ref{fig:RTsimul_Vf_wsp} we show our simulation results for those bi-metals at room temperature and with average layer thicknesses of the first metal of 100nm as a function of volume fraction $V_f$ of the second metal (which in most of our examples is the less conductive metal).
The second metal's layer thickness equals that of the first metal only at $V_f=0.5$ and is typically thinner when the second metal's volume fraction $V_f<0.5$ (see e.g. \cite{Ding:2021,Carpenter:2023a}).
Note that in cases where the second metal is significantly less conductive than the first, the volume fraction dependence is almost linear.
The only exception here is Cu/Ag because both metals are good conductors in this case.
%% previous 100T design: \cite{Freudenberger:2002} cites 60% IACS and 1.2 GPa UTS (below is for up to 120T LANL design)
Magnetic fields of about 100T are achievable with bi-metals with an electric conductivity of 60\% IACS together with an ultimate tensile strength (UTS) of 1.2 GPa \cite{Freudenberger:2002}.
In order to push the highest achievable magnetic fields well beyond 100T (say up to 120T), the UTS needs to be even higher \cite{Dubois:2012,NRC2013}, i.e. 1.5 GPa, without reducing electric conductivity below 55\% IACS;
this limit is indicated as a gray dashed line in Fig. \ref{fig:RTsimul_Vf_wsp}.
For most of the listed bi-metals, this means a volume fraction not much higher than 1/3 should be considered.

We therefore focus on $V_f\approx0.3$ for Figure \ref{fig:Vf03simul_T_wsp} where we show the temperature dependence of electric conductivity for all bi-metals considered in this work.
55\% IACS is again indicated in a dashed gray line, showing that this lower limit is exceeded even for the least conductive bi-metal, Cu/Fe, up to elevated temperatures of almost 350K.
We see that the temperature dependence in the range we simulated, i.e. 100--450\,K (in 50\,K increments), is very similar across all simulated bi-metals.

\section{Review of strength properties}
\label{sec:strength}
%%%%%%%%%%%%%%%%%%%%%%%%%%%%%%%%%%%%%%%

% \begin{itemize}
% \item Cu/Nb ARB \cite{Ding:2021}, $V_f=1/3$: UTS 1.2 GPa, $\sim61$\% IACS 
% \item Cu/Nb wire composites (ADB) \cite{Bevk:1978}: UTS up to 1.4GPa for $V_f=14.8\%$ and up to 2.2 GPa for $V_f=18.2\%$
% \item Cu/Nb, ADB \cite{Rozhnov:2019}: 1125 MPa UTS (and 68\% IACS) for 18 wt. \% Nb
% \item Cu/Nb, ADB \cite{Dupouy:1995}: UTS 1050 MPa for $V_f=25\%$
% \item Cu/Nb, hot pressed (HP) \cite{Lei:2013}: 1102 MPa UTS and 57\% IACS for 10 wt. \% Nb
% \item Cu/Fe, ADB \cite{Dupouy:1995}: UTS 1370 MPa for $V_f=0.5$
% \item Cu/Ta, cross accumulative roll bonding (CARB) \cite{Zeng:2016}, $V_f=1/3$: UTS up to 950 MPa for nominal (=Ta) layer thickness 50nm (implies Cu layer thickness 100nm), and UTS up to 800 MPa for nominal layer thickness 100nm
% \item Cu/Ag, CARB \cite{You:2021}, $V_f=1/3$: UTS $\sim680$ MPa at 100nm layer thickness, $\sim840$ MPa at 50nm layer thickness, and 940 MPa at 20nm layer thickness
% \item Cu/Ag, ADB \cite{Hong:1999}: up to $\sim1.03$ GPa UTS at 24 wt. \% Ag
% \item Cu/Cr alloys, high pressure torsion (HPT) \cite{Dobatkin:2015}: microhardness (HV) up to 1.9 GPa for $V_f=9.85\%$ (with 76 \% IACS and grain size of 143nm)
% \item Cu/W, copper-coated tungsten, no sever plastic deformation \cite{Han:2022}: up to >900 MPa at 60 wt. \% W
% \item Cu/Fe, ADB \cite{Yang:2022}: 0.863 GPa UTS for $V_f=0.2$ with 47\% IACS
% \item Cu/Ag, ADB \cite{Sakai:1991}: 1 GPa UTS and 80\% IACS for 16 atomic \% Ag
% \end{itemize}

\begin{figure}[!htb]
\centering
\includegraphics[width=\textwidth]{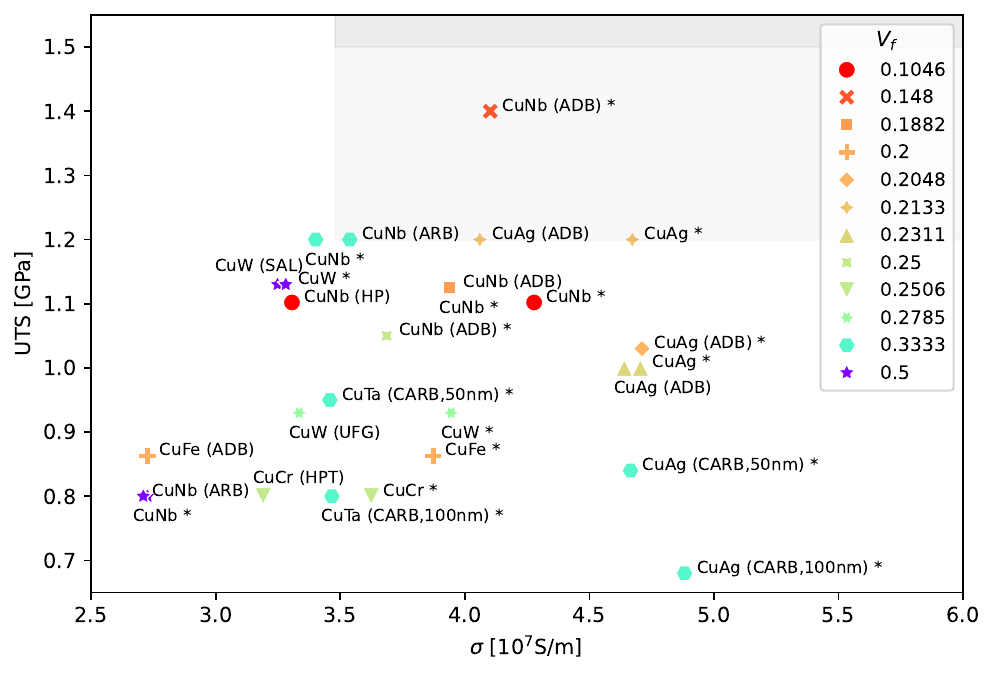}
\caption{In this overview over strength and electric conductivity of various bi-metals,
the shaded area indicates the requirements for generating magnetic fields of 100T and beyond.
The material and processing method are indicated in the marker annotations, where ADB refers to `accumulative drawing and bonding', ARB denotes `accumulative roll bonding', CARB refers to `cross accumulative roll bonding', UFG denotes `ultrafine-grained', HP refers to `hot pressed', HPT is `high-pressure torsion', and `SAL' is `self-assembled lamellar architecture'.
The volume fraction of the second metal is indicated in the legend.
For some roll bonded bi-metals, we also indicate the `nominal' layer thickness, i.e. at a volume fraction of 1/3 the nominal layer thickness is that of the second metal and the Cu layer thickness is twice that number (or rather, two Cu layers are always bonded together).
In cases where the electric conductivity was not reported, we use our own simulation result for the purpose of this plot, and indicate so by a star $*$.
Additionally, we compare our simulation results (for 100nm Cu layers within ARB) to all cases where conductivity was reported in the literature (indicated by a star $*$ once more).
Experimental data are taken from Refs.
\cite{Bevk:1978,Dupouy:1995,Lei:2013,Rozhnov:2019,Ding:2021,Carpenter:2022} (Cu/Nb), \cite{Dupouy:1995,Yang:2022} (Cu/Fe), \cite{Zeng:2016} (Cu/Ta) 
% \cite{Dobatkin:2015} (Cu/Cr),
\cite{Mao:2022,Prosviryakov:2016} (Cu/Cr),
\cite{Sakai:1991,Hong:1999,Grunberger:2002,Freudenberger:2002,You:2021} (Cu/Ag)
\cite{Han:2022,Han:2024} (Cu/W).
% For comparison, data points marked by a star $*$ correspond to our own simulation results for electric conductivity combined with the strength value reported in the literature.
}
% \cite{Ding:2021,Bevk:1978,Rozhnov:2019,Dupouy:1995,Lei:2013,Zeng:2016,You:2021,Hong:1999,Dobatkin:2015,Han:2022,Yang:2022,Sakai:1991}.}
\label{fig:conduct_vs_strength_overview}
\end{figure}

A number of bi-metals processed with various techniques have been studied with respect to their strength properties in the past.
ADB Cu/Nb wires have been of particular interest with ultimate tensile strengths up to well beyond 1 GPa, see Refs. \cite{Bevk:1978,Rozhnov:2019,Dupouy:1995}.
More recently, ARB as well as CARB (i.e. cross accumulative roll bonding, where every rolling step is undertaken perpendicular to the previous one) have been studied as well for a number of bi-metals including Cu/Nb, Cu/Ag, and Cu/Ta (among others).
Strengths have been achieved up to 1.2 GPa \cite{Ding:2021}, see also Refs. \cite{Zeng:2016,You:2021}.
Other processing techniques have been used as well, such as high pressure torsion (HPT) for e.g. Cu/Cr \cite{Dobatkin:2015}.
Apart from severe plastic deformation, researchers have also studied creating very strong bi-metals by reducing the grain sizes; for example the authors of Ref. \cite{Han:2022} achieved strengths above 0.9 GPa with ultrafine grained Cu/W.
Some of the studies mentioned above also measured electric conductivity of their bi-metal, such as e.g. Ref. \cite{Ding:2021} who achieved $\sim61$\% IACS for ARB Cu/Nb, Ref. \cite{Dobatkin:2015} who achieved 76 \% IACS for HPT Cu/Cr, and Refs. \cite{Carpenter:2022,Carpenter:2023a,Han:2024}.
Many others have focused solely on strength properties.

In Figure \ref{fig:conduct_vs_strength_overview} we combine strength and conductivity data from the literature with our own conductivity simulation results.
%where no conductivity was provided in  the reference.
The material and processing methods of the experimental data from the literature are indicated in the marker annotations, where ADB refers to `accumulative drawing and bonding', ARB denotes `accumulative roll bonding', CARB refers to `cross accumulative roll bonding', UFG denotes `ultrafine grained', HP refers to `hot pressed', HPT is `high pressure torsion', and `SAL' is `self-assembled lamellar architecture'.
In all those cases where the electric conductivity was not reported, we use our own simulation result for the purpose of this plot, and indicate so by a star $*$.
Additionally, we compare our simulation results to all cases where conductivity was reported in the literature (indicated by a star $*$ once more).

Note that our simulations pertain to ARB materials, though we may expect those results to be close enough to the conductivity of ADB bi-metals, provided we simulated for layer thicknesses well above the electron mean free path for the (more conductive) matrix material.
For those data where we have conductivity results from the literature, we compare to our own predictions and see that (as expected) they are very close if the bi-metal was processed with ARB, most notably the conductivity reported by Ding et al. \cite{Ding:2021}.
As for ADB processed materials, as strength is increased through additional drawing and bonding steps, conductivity decreases, and therefore our ARB-optimized simulations overpredict conductivity for the ADB material (see Figure \ref{fig:conduct_vs_strength_overview}) since we do not account for the actual ADB microstructure in these cases.
For this reason, the conductivity prediction for ADB Cu/Nb with UTS around 1.4 GPa shown in this figure is to be taken with a grain of salt: conductivity is likely somewhat lower than predicted.

For the purpose of Fig. \ref{fig:conduct_vs_strength_overview}, we assumed 100nm average layer thickness for the Cu phase in our simulations unless stated otherwise.
Those results are meant to give readers an overview over bi-metals with promising strength and conductivity properties that are worth being studied further.

The main reason to study ARB instead of ADB is that ADB materials, do not necessarily have the desired shape and have been unreliable despite their initially good properties, i.e. fractured filaments are introduced during the preparation process \cite{Ding:2021,Carpenter:2022}.
As we see from Figure \ref{fig:conduct_vs_strength_overview}, ARB materials have not achieved quite as high strength compared to ADB, but ARB has not been studied as long as ADB and researchers are constantly improving the former \cite{Ghalehbandi:2019,Ding:2021,Carpenter:2023a}.

From Fig. \ref{fig:RTsimul_Vf_wsp}, we see that a volume fraction of 1/3 or less for the less conductive material is necessary to achieve electric conductivities well above 55\% IACS.
Not surprisingly, many authors have thus focused on $V_f=1/3$, as shown in Fig. \ref{fig:conduct_vs_strength_overview}.
Clearly, ARB materials need to become even stronger to meet the 1.5 GPa UTS requirement for the next generation magnets \cite{Dubois:2012,NRC2013}.
%In order to reach this goal, (one of) the currently most promising candidates, Cu/Ag, CuW, and Cu/Nb, would need to be made even
%stronger than the current highest results of 1.2 GPa reported by Ding et al. \cite{Ding:2021} for Cu/Nb 
Our results and those from the literature indicate that Cu/Ag, CuW, and Cu/Nb exhibit conductivity and strength metrics that are closest to those needed for greater than 100T magnets.
However a strength greater than 1.2 GPa (reported by Ding et al. \cite{Ding:2021} for Cu/Nb) still needs to be achieved
while maintaining comparable electric conductivity (i.e. in order to move from the light-shaded into the dark-shaded area within Fig. \ref{fig:conduct_vs_strength_overview}).

\section{Conclusion and Outlook}
\label{sec:con}
%%%%%%%%%%%%%%%%%%%%%%%%

Ultrastrong magnetic fields require materials with high electric conductivity as well as high UTS to withstand the necessarily strong Lorentz forces.
Pushing the limits beyond 100T has been a challenge and in order to aid with the identification of next generation materials for these applications, we calculated from theory electric conductivities for a number of bi-metals processed by ARB.
Furthermore, we presented a survey of previous strength and conductivity results from the literature.
The latter conductivities were subsequently compared to our predictions (see Fig. \ref{fig:conduct_vs_strength_overview}), which are fairly accurate for ARB, but (not surprisingly) often overpredict conductivity for ADB and otherwise processed bi-metals.
Adapting our underlying model of microstructure together with accurate measurements of the actual microstructure will no doubt lead to more accurate predictions.

\subsection*{Acknowledgments}
%%%%%%%%%%%%%%%%%%%%%%%%%%%%%%%%%%%%%%%%%%

This work was supported by the U.S. Department of Energy through Los Alamos National Laboratory,
which is operated by Triad National Security, LLC, for the National Nuclear Security Administration of the U.S. Department of Energy under contract 89233218CNA000001.
This work was funded through Los Alamos National Laboratory Directed Research and Development (LDRD) project  ER20200375.

%%%%%%%%%%%%%%%%%%%%%%%%%%%%%%%%%%%%%%%%%%%
\bibliographystyle{utphys-custom}
\bibliography{microstructure}
\end{document}